\newcommand{\del}{\partial}

\newcommand{\extd}{{\rm d}}
\newcommand{\R}{\mathbb{R}}
\newcommand{\C}{\mathbb{C}}

\newcommand{\eps}{{\epsilon}}
\newcommand{\tens}{\mathop{\otimes}}

\newcommand{\note}[1]{}

\newcommand{\lcross}{{>\!\!\!\triangleleft}}
\newcommand{\cobicross}{{\triangleright\!\!\!\blacktriangleleft}}

\newcommand{\cp}{\mathfrak{p}}

\newcommand{\eqn}[2]{\begin{equation}#2\label{#1}\end{equation}}

\documentclass[11pt]{article}
\usepackage{amssymb,amsmath,amscd}

\allowdisplaybreaks 

\begin{document}


\begin{flushright}
CERN-TH/99-213\\
DAMTP/99-83\\
hep-th/9907110 \\
$~$ \\
July 1999
\end{flushright}

\medskip

\begin{center}

{\Large WAVES ON NONCOMMUTATIVE SPACETIME \\
AND GAMMA-RAY BURSTS}
\\ \baselineskip 13pt{\ }\\
\medskip
\medskip
{\ }\\ {\bf Giovanni AMELINO-CAMELIA}\footnote{Marie Curie
Fellow (permanent address: Dipartimento di Fisica, Universit\'a di Roma
``La Sapienza'', Piazzale Moro 2, Roma, Italy)}\\ \smallskip
Theory Division, CERN, CH-1211, Geneva, Switzerland
\\
\medskip
\medskip
$~$\\
{\bf Shahn MAJID}\footnote{Royal Society University Research Fellow and
Fellow of Pembroke College, Cambridge}\\ \smallskip Department of
Applied Mathematics \& Theoretical Physics\\ University of
Cambridge, Cambridge CB3 9EW\\
\end{center}
\medskip
\medskip

\begin{quote}
\noindent{\bf Abstract}
Quantum group Fourier transform methods
are applied to the study of processes
on noncommutative Minkowski spacetime $[x^i,t]=\imath\lambda x^i$.
A natural wave equation is derived and the associated
phenomena of {\it in vacuo} dispersion are discussed.
Assuming the deformation scale $\lambda$ is of the order of
the Planck length one finds that the
dispersion effects are large enough
to be tested in experimental investigations
of astrophysical phenomena such as gamma-ray bursts.
We also outline a new approach to the
construction of field theories on the noncommutative spacetime, with
the noncommutativity equivalent under Fourier transform to
non-Abelianness of the `addition law' for momentum in Feynman diagrams.
We argue that CPT violation effects of the type testable using the
sensitive neutral-kaon system are to be expected in such a theory.

\end{quote}

\baselineskip 18pt

\section{Introduction}

Quantum groups and their associated noncommutative geometry have
been proposed as a candidate for the generalisation of geometry
needed for Planck scale physics in \cite{Ma:pla,Ma:the}. Using
such methods there were provided models exhibiting the unification
of quantum and gravity-like effects into a single system with a
flat space quantum limit when a parameter ${\rm \scriptstyle G}\to
0$ and a classical but curved space limit when $\hbar\to 0$.
Radically new phenomena at the Planck scale were also proposed,
notably an extension of wave-particle duality between position and
momentum via Fourier theory to a novel duality between quantum
observables and states in which their roles could be interchanged.
For the specific models in \cite{Ma:pla,Ma:the} this
observable-state duality was implemented through quantum group
duality with the dual system of the same form but with different
values of the parameters (i.e. a form of T-duality). In addition,
in a different approach, it was proposed in \cite{Ma:reg} that
noncommutative geometry in the form of $q$-deformation could
provide an effective way to model Planck-scale quantum corrections
to spacetime geometry. It was argued that field theory on such
spacetimes would be more regular, with UV divergences appearing as
poles at $q=1$, while symmetries would be preserved as quantum
group symmetries. More recently in Refs.~\cite{gacxt,gacgrf98} an
analysis of some candidate quantum-gravity phenomena was used to
suggest that an effective large-distance description of some
aspects of quantum gravity might be based on quantum symmetries
and noncommutative geometry, while it was argued that at the
Planck scale even more novel structures might be required. In
particular, it was observed that the ``classical-apparatus
limit'', which is fully consistent~\cite{rose,wign} with ordinary
Quantum Mechanics, is not accessible~\cite{gacmpla} in theories
with gravitation, and this was used to suggest~\cite{gacgrf98}
that a fully developed quantum gravity should be based on a
mechanics departing from ordinary quantum mechanics in such a way
as to accommodate a new concept of apparatus and an accordingly
modified relation between the apparatus and the system under
observation.

In the present article we report progress toward the use of some
of these ideas in a testable and workable approach to particle
physics on noncommutative spacetime.

Since Planck-scale energies are so very far from present-day
experiments we will be mostly attempting to model
quantum gravity effects at distances much larger than the
Planck length, postponing to future work the investigation of
whether quantum-group ideas might prove useful for a description
of physics at even shorter distances (perhaps all the way down to
the Planck-length).
Specifically, we are interested in noncommutative Minkowski space as a
basis for an effective description of phenomena associated to a
nontrivial ``foamy'' quantum gravity vacuum of the type considered
by Hawking, Wheeler and others.
When probed very softly such a space would appear as an ordinary Minkowski
space, but probes of sufficiently high energy would be affected by the
properties of the quantum-gravity foam and we attempt to model
(at least some aspects of) the corresponding dynamics using a
noncommutative Minkowski spacetime.
In the present work we do not discuss the generalization necessary for
a description of how the quantum-gravity foam affects spaces which are
curved (non-Minkowski) at the classical level.
Even for spaces which are Minkowski at the classical level
a full quantum gravity of course would predict
phenomena which could not be simply encoded in noncommutativity of
Minkowski space and actually would not be exclusively associated
to its foamy vacuum structure, but it is plausible that the most
significant implications of quantum gravity for the low-energy
(large-distance) physics of Minkowski spaces
would be associated to some aspects of
the Hawking-Wheeler foam.

The noncommutative Minkowski spacetime
we consider here is the algebra
\eqn{mink}{ [x^i,t]=\imath\lambda \, x^i,\quad [x^i,x^j]=0}
where $i,j=1,2,3$ and $\lambda$ is a free length scale, which (as
justified by the above physical motivation for our studies) we
shall often implicitly assume to be closely related to the Planck
length $L_p \sim 10^{-35}m$. We work in units such that the
speed-of-light constant is $c=1$. The algebra (\ref{mink}) can be
interpreted as a version of Minkowski spacetime with
noncommutative coordinates, see notably
\cite{MaRue:bic}.
Such algebras in 3 dimensions can be found in
\cite{Ma:the} while a q-deformation version of them in 1+1
dimensions was further studied from a noncommutative spacetime
 point of view in \cite{Ma:reg}.
They provide a compelling candidate for the type of spacetime
in which we
are physically interested because they have a natural interpretation
in momentum space\cite{MaOec:twi}
and because they fit well the intuition emerging from certain
heuristic analyses of the structure of the quantum gravity
vacuum \cite{gacxt,gacmpla}. They are also part of a
2-parameter family of algebras proposed for Planck-scale physics
in \cite{Ma:pla}.

Among the already-studied implications of adopting (\ref{mink}) is
that the appropriate notion of Poincar\'e invariance under which it
is covariant has to be modified and becomes in fact a quantum
group\cite{MaRue:bic} (using the notation $P$ for momentum space)
\eqn{poin}{U(so_{1,3})\cobicross \C(P)}
of the bicrossproduct type introduced in \cite{Ma:the} in the
3-dimensional Euclidean case.  The paper \cite{MaRue:bic} also
showed that (\ref{poin}) was (nontrivially) isomorphic to the
so-called $\kappa$-deformed
Poincar\'e quantum group \cite{LNRT:def},
which had been earlier introduced
from another  point of view.
Ref.~\cite{MaRue:bic}
identified (\ref{mink}) as the spacetime on which
$\kappa$-Poincar\'e acts covariantly.
The introduction of a noncommutative-geometric point of view in which
the $\kappa$-Poincar\'e indeed acts covariantly on a suitable
 $\kappa$-Minkowski spacetime (\ref{mink}) was the main
result in \cite{MaRue:bic}. The paper also solved the problem
of finding the coordinate
algebra dual to the $\kappa$-Poincar\'e algebra and allowed it to
be identified it with an otherwise unconnected proposal in \cite{Zack}.

Preliminary, but to some extent heuristic, analyses of the
physical implications of the deformed or $\kappa$-Poincar\'e
proposal have led to interesting hypotheses, most notably the
possibility of modified dispersion relations \cite{kpoin,aln}.
Since recent progress in the phenomenology of gamma-ray bursts
\cite{grbnew} and other astrophysical phenomena
\cite{billetal,aemns2} renders experimentally accessible
\cite{grbgac} such modified dispersion relations, there is strong
motivation in verifying whether the analyses reported in
\cite{kpoin,aln} can be seen as part of a wider systematic
analysis of particle-physics phenomena in the noncommutative
Minkowski spacetime (\ref{mink}). This is one of the primary
objectives of the present paper. We shall rely on a different
approach suggested by \cite{Ma:ista} that makes use more directly
of the structure\cite{MaRue:bic} of (\ref{mink}) itself as a
quantum group in its own right. Its coproduct structure here is
\eqn{coprod}{ \Delta t=t\tens 1+1\tens t,\quad \Delta x^i
=x^i\tens 1+1\tens x^i}
which expresses the addition law on (\ref{mink}). As noted already
in \cite{MaRue:bic}, this addition law is valid but is not itself
covariant under the deformed Poincar\'e algebra, hence this
algebra necessarily takes a back seat in the new approach. We deal
separately with translation and (classical) Lorentz covariance.
However, this new approach based directly on the intrinsic
structure of (\ref{mink}) allows us to make substantial progress
toward a formalism in which all computations are not significantly
more complicated than in a corresponding ordinary theory in a
commutative spacetime. Important tools are provided by the
availability of a 4-dimensional translation-invariant differential
calculus\cite{Oec:cla} (which is not possible in the
$\kappa$-Poincar\'e covariant setting) and by the quantum group
Fourier transform which was worked out for our particular algebra
in \cite{MaOec:twi}. The latter is
defined by the additive quantum group structure and allows one to
rewrite structures living on noncommutative spacetime as
structures living on a commutative (but nonabelian)
``energy-momentum'' space. Since our emphasis is on the structure
of the space-time (\ref{mink}) we find it convenient to write
formulas in terms of the length scale $\lambda$ rather than the
dimensionful parameter $\kappa$ of the $\kappa$-Poincar\'e
approach. Because of the transparent 
relation $\lambda =\hbar \kappa^{-1}$ we 
do not expect our choice of conventions to
create any confusion; however, for good measure, we shall
occasionally refer back to the ``$\kappa$'' notation and emphasize
that some of the structures we consider are frequently denominated
in the literature as $\kappa$-Minkowski spacetime and
$\kappa$-Poincar\'e group.

In the next Section we
provide the basic elements of the mathematics
used for our proposal:
we discuss the analogues of functions and a
differential calculus on the deformed Minkowski
spacetime (\ref{mink}) and then use the above-mentioned Fourier transform
to introduce the deformed momentum space and
the wave equation
for deformed Minkowski spacetime.
In Section~3 we discuss the physical interpretation
of some of the new structures present in
deformed Minkowski spacetime.
In Section~4 we
provide a possible scenario in which (\ref{mink}) arises as
the quantum system associated to spacetime itself as the phase
space of some `pregeometry'.
In Section~5, also using the Fourier theory,
we sketch out a proposal for a new approach
to the construction of field theories
in noncommutative spacetimes.
In Section~6 we elaborate on the phenomenology
associated to {\it in vacuo} dispersion and
CPT violation.
Finally in Section 7 we summarize our results and
set up an agenda for future work.

\section{Functions, Differential Calculus, Momentum space
and Wave Equation}

A key ingredient of our proposal (and a general feature of the
particular quantum groups in \cite{Ma:pla}) is that all functions
in $x^i,t$ can be treated as if classical under a normal ordering
prescription. Thus, we consider general functions on the deformed
Minkowski spacetime (\ref{mink}) as elements of the algebra of the
form $:\psi(\vec x,t):$ where $\psi$ is a usual function in 4
variables and where by convention the $t$ generator is taken to
the right.

The translation coproduct (\ref{coprod}) implies a natural
4-dimensional translation-invariant calculi of
differentials\cite{Oec:cla} spanned by $\extd x^i,\extd t$. In
noncommutative geometry the differential
calculi are not usually unique but in the 1+1 dimensional case of
(\ref{mink})
there are in fact two possibilities discussed in \cite{Oec:cla}; we
chose one of these to extend to our 4-dimensional case, namely with the
relations
\eqn{extd1}{ [x^i,\extd x^j]=0,\quad [t,\extd x^i]=0,\quad
[x^\mu,\extd t]={\imath \lambda } \extd x^\mu.}
The corresponding partial
derivatives are defined by
\eqn{partial}{ \extd \psi= (\del_\mu\psi)\extd x^\mu}
and take the form
\eqn{del1}{\del_i:\psi(\vec{x},t):=:\frac{\del}{\del x^i}\psi(\vec{x},t):
,\quad \del_0:\psi(\vec{x},t):=
:(\imath\lambda)^{-1}\left(\psi(\vec{x},t)
-\psi(\vec{x},t-\imath\lambda)\right):}
This means that the associated noncommutative differential
geometry of our deformed Minkowski space behaves in practice like
the usual differentials in position and like a lattice in the time
direction. On the other hand, $t$ is an operator and there is no
fixed lattice in this noncommutative geometry. Concerning the
nature of the time-direction lattice it would be tempting here to
redefine, say, $i\lambda=\mu$ so that $\del_0$ appears like a
usual lattice derivative. However, if $\mu$ is real then
(\ref{mink}) tell us that for hermitian $x$, $t$ would have to be
antihermitian. So in any conventional ideas of measurement it
would have imaginary eigenvalues. One would then be displacing
imaginary time values by real $\mu$. Since we prefer to envisage
real eigenvalues for $t$ we are forced to take $\lambda$ real. In
fact the ``$\imath$'' here is not so alarming. On analytic
functions we obtain as $\lambda \rightarrow 0$ the usual
differential just as well as for a real displacement, so this is
an equally valid deformation even if a little unfamiliar. In fact
its meaning is that this $\del_0$ {\em is a lattice differential
in Euclidean space} and just appears as above after Wick rotation.
We recall that frequently in theoretical physics certain
constructions look more natural in Euclidean space and are only
viewed in Minkowski space after Wick rotation. This would appear
to be such a situation.

Next we consider integration in our non-commutative spacetime.
A natural translation-invariant choice is
\cite{Ma:reg}
\eqn{int}{ \int :\psi:=\int \extd^3\vec{x}\ \extd t\ \psi(\vec{x},t)}
in terms of usual integration of the underlying function. It is
such that the integral of a partial derivative of a
suitably decaying function $\psi$ vanishes.

We now consider the momentum space dual under
non-Abelian Fourier transform
to the Minkowski spacetime (\ref{mink}).
Note first of all that Fourier
theory is usually considered for Abelian
groups but the nonAbelian case
can be handled just as well using modern
(quantum group) methods. Thus, if $P$ is some nonAbelian matrix group then its
algebra of coordinate functions $\C(P)$ can be regarded as a (commutative)
quantum group or Hopf algebra.
Its dual (cocommutative) Hopf algebra is the enveloping algebra
$U({\cp})$ where $\cp$ is the Lie algebra of $P$ and Fourier transform
provides maps $\C(P)\to U({\cp})$ and vice-versa. This is a completely
canonical construction\cite{Ma:book}, but it does oblige us to
regard the enveloping algebra $U({\cp})$
as the `coordinates' of some noncommutative space if we want to think
of Fourier theory as mapping functions on one space to `functions' on some
dual space (this is why usual Fourier
theory is restricted to Abelian groups so that the dual is a usual and not
noncommutative space).

Our Minkowski spacetime (\ref{mink}),
(\ref{coprod})
is such an enveloping algebra and is therefore connected by nonAbelian
Fourier theory precisely to functions on a classical but
nonAbelian momentum group, namely the group $P$ of matrices of the
form
\eqn{Pmat}{
 \begin{pmatrix} e^{\lambda\omega} & k_1& k_2& k_3
\\ 0 & 1&0&0 \\ 0&0&1&0\\
 0&0&0&1 \end{pmatrix}.}

One may then compute the canonical Fourier theory using the integral
(\ref{int}) and the canonical element for the duality pairing\cite{Ma:book}.
It comes out as cf.\cite{MaOec:twi}
\eqn{fou}{ T:{\rm Mink}_\lambda\to \C(P),\quad T(:\psi:)
(\vec{k},\omega)=
\int \extd^3\vec{x}\ \extd t\  e^{\imath\vec{k}\cdot\vec{x}}e^{\imath \omega t}
\psi(e^{\lambda\omega}\vec{x},t)}
where the integral is over usual commuting functions. The canonical
property of Fourier theory comes out as
\eqn{foudel}{ T(\del_i\psi)=-T(\psi)\imath k_ie^{-\lambda\omega},\quad
T(\del_0\psi)=-T(\psi) \, \imath \,
\frac{1-e^{-\lambda\omega}}{\lambda} \, .}
We can also work with the generator of $\del_0$ as
\eqn{foudelt}{\del_t:\psi:=:\frac{\del}{\del t}\psi:,\quad T(\del_t\psi)
=-T(\psi)\imath\omega.}
These
formulae emerge naturally from noncommutative geometry and of
course become usual Fourier theory when $\lambda\to 0$. Here
\eqn{group}{(\vec{k},\omega)(\vec{k}',\omega')=
(\vec{k}+e^{\lambda\omega}\vec{k}',\omega+\omega'),
\quad (\vec{k},\omega)^{-1}=(-\vec{k}e^{-\lambda\omega},-\omega)}
are the group law and inversion in the nonAbelian momentum group.

The natural plane waves associated to a point $(\vec{k},\omega)$
in momentum space are provided by the inverse Fourier transform of
left-translation invariant delta-functions at
$(\vec{k},\omega)^{-1}$, which come out as
\eqn{wave}{\psi_{\vec{k},\omega}=e^{\imath\vec{k}\cdot \vec{x}}
e^{\imath\omega t},}
i.e. a plane wave in our deformed Minkowski spacetime. These respect
the group law on momentum space in the sense
\eqn{prodwaves}{\psi_{(\vec k,\omega)(\vec k',\omega')}= \psi_{\vec k,\omega}
\psi_{\vec k',\omega'}}
so that, in particular, the wave in the reverse direction in
momentum space is
\eqn{revwave}{\psi_{(\vec{k},\omega)^{-1}}=
e^{-\imath\vec{k}e^{-\lambda\omega}\cdot\vec{x}}e^{-\imath\omega t}
=e^{-\imath\omega t}e^{-\imath\vec{k}\cdot\vec{x}},}
i.e., another
plane wave in our deformed Minkowski space time (note, however,
the order of the generators.)

We are now ready to obtain
the appropriate dispersion relations for such
waves. By definition these are constraints in momentum space $P$
which should be Lorentz invariant. Because our momentum space is
a nonAbelian group, not the usual $\R^{1,3}$, the appropriate action
of the Lorentz algebra is not the usual one. Rather, there is
a particular action of the Lorentz algebra on the momentum group
$P$ which is used in the semidirect product algebra
(\ref{poin}) of the deformed Poincar\'e quantum group in
\cite{MaRue:bic}. We clearly should use this action. It is \cite{MaRue:bic}
\eqn{loract}{M_i=-\eps_{imn}k_m\frac{\del}{\del k_n},\quad N_i
=k_i\frac{\del}{\del \omega}-(\frac{\lambda}{ 2}\vec{k}^2+
\frac{(1-e^{2\lambda\omega})}{2\lambda})\frac{\del}{\del k_i}
+\lambda k_ik_j\frac{\del}{\del k_j}} for
the action of the standard rotation and boost generators. These
are the vector fields on $P$ corresponding to the action on
generators given in \cite{MaRue:bic}.

{}From (\ref{loract}) one finds
the appropriate constraint which has the right limit and
which is both Lorentz invariant and
invariant under group inversion (\ref{group}) to be
\eqn{disp}{\lambda^{-2}\left(e^{\lambda\omega}
+e^{-\lambda\omega}-2\right)-\vec{k}^2e^{-\lambda\omega}
=m^2.}
The operator corresponding under Fourier theory (\ref{foudel}) to
the left hand side in momentum space is $-\square$, where
\eqn{cas}{\square:\psi:=:-\lambda^{-2}\left(\psi(\vec{x},t+\imath\lambda)
+\psi(\vec{x},t-\imath\lambda)-2\psi(\vec{x},t)\right)-\sum \del_i^2
\psi(\vec{x},t+\imath\lambda):}
i.e.,
\[  \square=(\del_0^2-\sum\del_i^2)L\]
where $L:\psi(\vec{x},t):=:\psi(\vec{x},t+\imath\lambda):$ is the shift
operator and $\del_0$ is the derivative in (\ref{del1}).
It is easy
to see that the plane waves (\ref{wave}) are eigenfunctions with
eigenvalue given by the left hand side of (\ref{disp}). Also, from
the bicrossproduct construction of the deformed Poincar\'e algebra
(\ref{poin}) in \cite{MaRue:bic} it is known that
a Lorentz-invariant expression in
momentum space necessarily corresponds to a Casimir from the
deformed Poincar\'e point of view.

It is of course important to be able to construct wave packets
from our plane-wave solutions. To construct a wave packet we
should average over waves with some density function $a$, for
example $a$ might be a Gaussian centred at the origin and then
translated to be centred at some average spatial momentum
$\vec{k}_0$ (and trivial in the energy direction). In more
conceptual terms the wave-packet is the inverse Fourier transform
of the translated $a$. In addition, the
composite waves are constrained to obey the dispersion relation.
The noncommutative analogue is therefore
\eqn{packet}{ \psi_{a,\vec k_0}(\vec x,t)
=\int \extd^3\vec k\ e^{\lambda \omega}
a((\vec k_0,\omega_0)(\vec k,\omega)^{-1})
\psi_{(\vec k,\omega)^{-1}}}
where $\omega$ is a function of $\vec k$ according to the
dispersion relation (\ref{disp}). Similarly for $\omega_0$. The
choice of $a$ a left-invariant delta function recovers a pure on
shell plane wave. With care one may also change the variable
$(\vec k,\omega)^{-1}$ of integration to $(\vec k,\omega)$.

Let us note that while this is the natural definition from the
mathematical point of view, the physical applications to which we
put our wave-packet might dictate other choices based on the same
pattern. Thus, in the above we have used the left-translation
invariant integral $\int\extd\omega\extd^3\vec k
e^{\lambda\omega}$ required
by the
quantum-group Fourier theory in the present conventions. It is
also possible that one might prefer to build a wave-packet using
an integration invariant under the deformed action of Lorentz
transformations. This would be with integration measure
\eqn{lorint}{\int \extd \omega ~  \extd^3 \vec{k} ~e^{-3\lambda \omega}}
so that $\int M_i(f)=0,\quad \int N_i(f)=0$ (where $M_i, N_i$ are
the rotation and boost vector fields) if $f$ is sufficiently
rapidly decaying at infinity. In other conventions or some other
applications one might also need right-invariant integration
measure
\eqn{rightint}{\int \extd \omega ~ \extd^3 \vec{k}}
so that $\int f_{(\vec{k}_0,\omega_0)}=\int f$, where
$f_{(\vec{k}_0,\omega_0)}
(\vec{k},\omega)=f((\vec{k},\omega)(\vec{k}_0,\omega_0))$. These
choices will be discussed further in Section~5.

\section{Physical Interpretation}

It is of course necessary
to discuss the relation between
the algebra of the deformed Minkowski spacetime (\ref{mink})
and the physically measured
time and position of events.
It is tempting to
associate to our formal normal ordering prescription
an operative prescription in which the
coordinates can be treated conventionally provided one
always measures the time coordinates first.
This is in fact what one would expect
based on an analogy with similar normal-ordering
prescriptions in ordinary quantum-mechanics frameworks.
While in the following we do
assume that there
exists some form of measurement procedure in an unknown theory
of quantum gravity
allowing us to treat our
coordinates conventionally, we want to emphasize that
the nature of the
observables associated to our operators must be
somewhat different from the observables of ordinary
quantum mechanics. We expect such differences
especially because we have a time operator,
while ordinary quantum mechanics only involves a time
parameter\footnote{The time appearing in the evolution equations
of ordinary quantum mechanics is indeed only a parameter. One can
attempt to construct in some way a ``time of arrival operator''
(see, {\it e.g.}, Ref.~\cite{rovtime}) but in
general there is no self-adjoint operator canonically conjugate
to the total energy, if the energy spectrum is bounded
from below~\cite{pauli}}.
The observables of ordinary quantum mechanics
are measured in correspondence with
a value of the time parameter, and at least
the observable associated to our time operator does not
appear suited for this type of operative definition.
Thus, one may attempt to treat the system with operators
$x_i,t$ quantum mechanically (we give an example in the next section) but
the time variable for that quantum mechanics would have to
 be different from the
operator $t$. The two times would at some point need to be reconciled
within a more complete and unknown theory of quantum gravity.
 In fact the problem we are facing
here is nothing else than another version
of the ``problem of time'' encountered in one form or another
in any approach to quantum gravity, although not always
immediately evident within some of the more abstract formalisms.

There is probably no reason to be surprised of these difficulties
of ordinary quantum mechanics.
In fact, the conceptual analysis of
measurements procedures~\cite{gacmpla,bergstac,ng}
for candidate quantum-gravity observables
has been used~\cite{gacgrf98,gacbig}
to argue that the mechanics on which quantum
gravity is based should not be exactly
the one of ordinary quantum mechanics. The new mechanics
should accommodate a somewhat different relationship between ``system''
and ``measuring apparatus'', and should take into account
the fact that the limit in which the apparatus behaves classically
is not accessible\footnote{In ordinary (non-gravitational)
quantum mechanics the limiting procedure allowing
to consider classical apparatus requires an infinite-mass
limit \cite{rose,wign}, which turns out to
be inconsistent with the structure of gravitational
measurements \cite{gacgrf98,gacmpla,bergstac,ng}.}
once gravitation is turned on.
The issue of separation between `observer' and `observed',
which is likely to play a central role in the new mechanics,
has already been explored to some extent
from the point of view of the necessary
formalisms in Refs.~\cite{Ma:pla} and in more recent works
such as \cite{rovsep}.
In general measurement is seen as
an interaction between aspects of the system labelled by macroscopic
or classical `handles' and the microscopic quantum system. To formulate
this properly one first needs to have a way to `identify' the macroscopic,
typically geometric, aspects (such as the separation
between two devices) within the overall quantum system, which is precisely
the task of noncommutative geometry.

For our present purposes, in trying to envisage a type of setup
that would allow to treat our
coordinates conventionally, let us consider
the measurement of the speed of a particle
travelling along a straight-line trajectory.
Assuming the space-time points
were identified in our laboratory by a grid of
clock-detector pairs
and that one was able to set up the emission of the
particle from position $P_0$ at time $T_0$,
the speed could be measured in two ways:
by measuring the clock time needed by the particle
to reach a given detector in the grid
or by measuring
the position (detector triggering) of the particle
at a given time of the (sychronized) clocks of our grid.
While any definite statement must await the development of a
consistent measurement theory
for quantum gravity (and in particular
for the class of models we are considering),
we expect that within our proposal these two ways of measuring
the speed would be significantly different and
it appears plausible that our normal ordering prescription
would correspond to the second method, the one in which
a chosen clock readout triggers the detectors to determine
the position of the particle at that time.

\section{An example of pregeometry quantum system}

While for the effective theory viewpoint that we advocate the
details of the underlying physics
are not directly
relevant, it might nevertheless be useful to have at
least an intuitive picture of the fact that our noncommutative
Minkowski spacetime should emerge from quantum gravity. We assume
that only certain macroscopic modes of the unknown quantum gravity
theory survive at the level of our effective description and, for
the sake of discussion, that these form an effective quantum
mechanical system underlying the $[x^i,t]$ noncommutativity
relations. We call this the {\em pregeometry quantum system}. It
should be considered as still an effective description of some
unknown quantum gravity theory but one which is slightly deeper
than the operative prescription for handling $x^i,t$ in terms of
classical functions in the preceding sections.

We should stress that our operative prescriptions for handling
$x,t$ as well as for scattering in terms of classical
momentum and
energy $\vec{k},\omega$ do not require us necessarily to develop this
extra layer of `pregeometry' for our model. Moreover, the best
description of the effective `pregeometry' may not be a quantum
one at all. Nevertheless, the conventional way of thinking about
noncommutative algebras is in terms of quantum mechanics and hence
it is natural to provide, for completeness, at least a sketch of
one example of a suitable quantum system that could serve as a
link between our operative description and the unknown
quantum gravity theory.

To approach this question, not knowing a complete quantum gravity
theory, we can nevertheless explore some mathematical
possibilities.
This is akin to using classical
topology to distinguish different {\em a priori} possible
classical solutions of a complex system, but in our case in an
algebraic or quantum mechanical setting. Thus we would like to ask
about different possibilities to extend the algebra to a quantum
system with additional $p_i$ generators and suitable commutation
relations between position and momenta subject to some {\em a
priori} assumptions.

This question was explored and answered in one spatial dimension
in \cite{Ma:pla}. Thus, if we are given a variable $x$ which we
deem to be position and a variable $\pi$ which we deem {\em a
priori} to be some kind of `momentum' variable and ask for {\em
all possible} commutation relations such that the addition law in
phase space $\R^2$ extends to the quantum system as a quantum
group $A$ extending $x$, $\pi$ in the sense
\[ \C[\pi]\to A\to \C[x]\]
as a Hopf algebra extension (here $\C[x]$ denotes functions in one
variable $x$, etc.) then one finds (coming out of the analysis) a
two-parameter family of possibilities\cite{Ma:pla} for $A$, namely
\eqn{planck}{ [x,\pi] = \imath \hbar_0 (1-e^{-\frac{x}{\rho}})}
where $\hbar_0,\rho$ are the two parameters, with the coproduct
\eqn{plankcop}{\Delta x=x\tens 1+1\tens x,\quad\pi=\pi\tens 1
+e^{-\frac{x}{\rho}}\tens \pi.} This is the 2-parameter
`Planck-scale Hopf algebra'
$\C[\pi]\cobicross_{\hbar_0,\rho}\C[x]$ introduced in
\cite{Ma:pla} in this way. Of course, that $A$ should {\em a
priori} be a Hopf algebra extension is a conceptual assumption
which may well not be true. I.e. we are not absolutely forced to
take this form of $A$, it is merely a mathematically natural class
of possibilities. Moreover, whereas in \cite{Ma:pla} the two
parameters were intepreted as the physical $\hbar$ and the
gravitational length scale of the background geometry, in our case
they are the parameters of the pregeometry system with an unknown
relationship to the actual physical parameters of the unknown
quantum gravity theory. Likewise, we do not suppose that $\pi$ is
exactly the physical momentum of the theory. Rather, we are merely
using the mathematical formalism of quantum mechanics to build a
deeper model behind the commutation relations (\ref{mink}).

In any event, motivated by this one-dimensional analysis, as an
example of a pregeometry quantum system for our 3+1-dimensional
Minkowski space we take three independent copies of
(\ref{planck}), i.e. we add generators $\pi_i$, say, where
$i=1,2,3$, with the relations and coproduct
\[ [x^i,\pi_j]=\delta^i{}_j
\imath\hbar_0(1-e^{-\frac{x^i}{\rho}}),\quad [\pi_i,\pi_j]=0,\]
\[ \Delta \vec{x}=\vec{x}\tens 1+1\tens\vec{x},\quad \Delta
\pi_i=\pi_i\tens 1+e^{-\frac{x^i}{\rho}}\tens \pi_i\]
Within this larger algebra we identify our noncommutative
Minkowski space as generated by the $x^i$ and
\[ t= \sum_i \pi_i\]
in the limit
\[ \rho,\hbar_0\to\infty,\quad \frac{\hbar_0}{\rho}=\lambda.\]
Note that the role of the $\pi_i$ here is as `auxiliary time
variables' with their sum
giving the time of the Minkowski theory. One in
fact expects something unusual like this when one considers the
asymmetric (and to date still problematic) treatment of time in
canonical quantum gravity; there one considers the spatial fields
and their conjugates on each time-slice and tries to reconstruct
the spacetime time afterwards. In addition, the commuting $\pi_i$
corresponds to the absence of spatial curvature in the
noncommutative Minkowski-space.
One can certainly envisage more
complex models
where an additional parameter enters into nontrivial
commutation relations between the $\pi_i$ as well.

Although this is just one example of a pregeometry quantum system,
it shows how the Minkowski space algebra (\ref{mink}) might arise
as the limiting case of commutation relations which have a more
familiar `quantum mechanical' form. (And if one wants to render
the commutation relations in an even more canonical form one need
only change to $\tilde\pi_i=(1-e^{-\frac{x^i}{\rho}})^{-1}\pi_i$
at the expense of a more unnatural coproduct in terms of
$x_i,\tilde\pi_i$.)

Given such a picture, one can now explore, at least tentatively,
certain issues. First of all, as a genuine quantum system in
\cite{Ma:pla} one has natural hermiticity properties
\[ x^i{}^*=x^i,\quad \pi_i^*=\pi_i\]
giving a Hopf $*$-algebra. We see that if we want to have a
quantum-mechanical interpretation of the noncommutativity of our
Minkowski-space then we should take $\lambda$ real when $t$ is hermitian.
This forces us to the imaginary finite differences in $\del_0$ in
Section~2. Alternatively we could 
replace $\imath\lambda$ by $\mu$ here and in Section~2 for a more 
conventional `lattice differential' in $\del_0$
but would then have to take $t$ antihermitian for a quantum mechanical
picture. This situation is not
unlike quantum field theory where for a proper mathematical
foundation it is best to Wick rotate to imaginary time. One might
therefore expect that this should be an effective remnant of the
problem of Wick rotation in the unknown quantum gravity theory.

One also has a natural `Schroedinger type' representation on
wavefunctions $\phi(\vec{x})$ with $x^i$ acting by multiplication
and $\pi_i=-\imath\hbar_0(1-e^{-\frac{x^i}{\rho}})\frac{\del}{\del
x^i}$, etc. The implied representation of our Minkowski space is
\eqn{rep}{x^i\cdot\phi=x^i\phi,\quad t\cdot\phi=-\imath\lambda\sum_i
x_i\frac{\del}{\del x^i}\phi} i.e., $t$ acts as an infinitesimal
scale transformation. The $x^i$ is hermitian and $t$ indeed
hermitian with respect to a certain inner product.

One has uncertainties in the simultaneous measurement of
$x^i,\pi_i$ and other familiar quantum effects. Of course it
implies the obvious uncertainty due to (\ref{mink}) but
potentially further uncertainties as well, depending on the
ultimate physical interpretation of the individual $\pi_i$.
Similarly, if one takes as in \cite{Ma:pla} the Hamiltonian
$\vec{\pi}^2/2m$, one has dynamics on the pregeometry quantum
system consisting of a particle moving more and more
slowly as it
approaches the origin (in a manner not unlike the approach to a
black hole event horizon\cite{Ma:pla}). Of course, the formal time
variable pertaining to this discussion of the pregeometry quantum system
should not be confused with the operator $t$
defined in (\ref{rep}) from the pregeometry momentum operators $\pi_i$.

We emphasize again that we are providing these comments solely for
illustrative purposes. Of course, experiments suitable for exploring the
nature of such a pregeometry system are well beyond our reach.
In principle one would first devise experiments to confirm
(or falsify) the models at the ``geometry level'' and only once
this level was well established one could hope to devise even more
refined experiments to test models of the ``pregeometry level''.
Since technology only very recently~\cite{grbgac,gacbig,gacgwi,elmn}
became advanced enough for a few very
preliminary experimental investigations of the ``geometry level'',
all considerations concerning the ``pregeometry level''
must indeed be considered as purely illustrative.

Finally, as well as the example discussed above based on
`extension theory' there are other more naive approaches to the
pregeometry quantum system one could also consider. For example,
for any Hopf algebra $H$ there is a canonical semidirect product
$H\lcross H^*$, the Weyl algebra or so-called Heisenberg double,
see \cite{Ma:book}. It is easy enough to compute in our case as
generated by $x^\mu,p_\mu$ with the commutation relations given in
\cite{MaRue:bic} as the action of the $p_\mu$ on the $x^\nu$ as
part of the action there of the deformed Poincar\'e quantum group.
While probably playing some role, we do not take it as the
pregeometry quantum system itself because as a `quantisation' it
treats time on the same footing as the space (which is not really
appropriate even when quantising a single relativistic particle).
The Weyl algebra also does not have a coproduct or other
interesting mathematical properties to characterise it in place of
that. We defer the discussion of this to further work in which,
particularly, the relationship between any pregeometry quantum
system and quantum mechanics {\em on} the noncommutative Minkowski
space (which are different questions) should be explored.

\section{Quantum field theory on noncommutative spacetime via
 nonAbelian energy-momentum space}

In this section we point out the possibility of a new approach to the
construction of field theories on a noncommutative space-time.
Previous attempts at a
satisfactory definition of field theory
in a non-commutative space-time
have had only limited success. At a rather formal level some
progress has been made,
but eventually one was confronted with
the difficulties involved in generalizing to a noncommutative
space-time some of the operators and other tools required
for a field theory.\footnote{In particular,
interesting studies of field theories in certain other
noncommutative geometries were reported in the
two preprints \cite{momspace} and \cite{oeckl}
which appeared while we were completing the writeup of the present article.
Our own approach is completely different from these works (in fact they are
based on different methods which would not seem to apply to our
particular spacetime (\ref{mink}) at all.)}
Here we observe that these difficulties could be evaded
by exploiting the fact that quantum group Fourier transform
allows us, as we have already seen, to rewrite structures living
on noncommutative spacetime as structures living on a classical
(but nonAbelian) ``energy-momentum'' space.
If one is content to evaluate everything in energy-momentum space,
this observation gives the opportunity to by-pass all problems directly
associated with the non-commutativity of space-time.
We are confident that eventually a compelling space-time formulation
of field theories on noncommutative geometries will emerge,
but in the meantime we restrict ourselves to energy-momentum space
where the underlying noncommutativity manifests itself only
through the ``curvature'' (nonabelianness of the group) of the space.
Note that this approach does not work for {\em any} noncommutative
spacetime but for all those where the spacetime coordinate algebra is
 the enveloping algebra of a Lie algebra, with the Lie algebra generators
regarded `up side down' as noncommuting coordinates\cite{Ma:ista}.

Because of the viewpoint we are advocating,
within our approach field theories are not naturally described
in terms of a Lagrangian. We resort directly to a Feynman-diagramatic
formulation.\footnote{This also implies that the description of
certain non-perturbative effects
(the ones not obtainable as infinite sums of Feynman diagrams)
might not be possible
within our energy-momentum space formulation.}
In principle, according to our proposal a given ordinary field theory
can be ``deformed'' into a counterpart living in a suitable noncommutative
spacetime not by fancy quantum group methods but simply
by the appropriate modification of the momentum-space
Feynman rules to those appropriate for a nonAbelian group.
The quantum group concepts are, however, required in order to do
this in a manner
consistent with the (noncommutative) geometry of space-time, for example to
 consistently obtain predictions for cross sections from the amplitudes
evaluated
using the nonAbelian Feynman rules.

While we postpone to future work (also because
some of the relevant mathematics is only at an early
stage of development) the detailed discussion of examples of such
field-theoretical models, in the rest of this section we give some
general guidelines to be followed in constructing the type
of field theories we are proposing.

Let us start with the Feynman rules. As mentioned the guiding
principle of our proposal for the construction of deformed field
theories is the replacement with quantum-group counterparts of
those group-theoretic elements which characterize the structures
relevant for ordinary field theories. Accordingly, the propagator
$D({\vec k}, \omega)$ of a scalar particle will be essentially
given by the inverse of the operator in the dispersion relation
Eq.~(\ref{disp}), i.e. in place of $D=(\omega^2 - {\vec k}^2
- m^2)^{-1}$ we take
\eqn{prop}{ D_\lambda= \left(\lambda^{-2}
(e^{\lambda \omega} + e^{- \lambda \omega} -2)
- e^{-\lambda \omega} {\vec k}^2
- m^2\right)^{-1}
~.}

The Feynman rules for vertices
that do not involve the momenta of incoming/outgoing
particles remain unchanged. For example, in ``$\Phi^4$'' theory
the 4-point vertex is still simply given by the coupling constant
\eqn{defvertnorm}{\Gamma = g \rightarrow \Gamma_\lambda = g ~.}

Vertices which involve the momenta of incoming/outgoing
particles and in particular those that require to sum the momenta
of pairs of particles must be rewritten
also taking into account the
rule (\ref{group}) for combining momenta in our deformed
Minkowski spacetime.
We postpone a full discussion
until we will be ready to discuss more complex field theories. It is
clear, however, that
when our momenta are nonAbelian there will be a fundamental difference
between
scattering particle 1 with particle 2 and scattering particle 2 with
particle 1; even for
trivial scattering the total momentum of particle 1 plus particle 2
or particle 2 plus particle 1 (where addition
is replaced by our nonAbelian group law) will be different in the two
cases. This is a new physical effect which we are predicting, which is
therefore difficult to lay down the rules for in advance. In the first
instance one should simply do scattering computations according to
all distinct order-of-addition rules, to see which fit best with a
given set of actual scattering experiments. One could also express
ignorance of the new effect by averaging over the different
orderings of the momenta.
Such an averaging procedure might even be the correct choice
at least in cases involving indistinguishable particles.

The Feynman diagrams involving integration over loop momenta will
also reflect the underlying non-commutativity of spacetime and
nonAbelian nature of energy-momentum space, through the measure of
integration on the latter. As mentioned in our discussion of wave
packets in Section~2, there are at least three candidates for the
measure of integration in energy-momentum space ({\it i.e.} for
loop integration); the left-invariant, right-invariant and Lorentz
invariant measures. All three coincide classically but in our
noncommutative theory we have to choose. Fortunately, all the
measures have a similar form
\eqn{alphaint}{\int \extd \omega ~ \extd^3 \vec{k} ~
e^{\alpha\lambda \omega}} for suitable
$\alpha=1,0,-3$. One can therefore proceed,
for example,
 with $\alpha$ regarded as a parameter to be fitted by comparison with
experminent.

Since Feynman rules come in fact from an analysis of the
scattering of wave-packets, the obvious choice suggested by
Section~2 is the left-invariant one
$\alpha=1$. However, we would prefer to leave the choice open at
the present stage. Future work might show that only some (perhaps
only one) of these candidates leads to renormalizable theories.
Actually, it is plausible that some of these measures might lead
to finite theories, since the exponential suppression of
high-energy modes might be sufficient to eliminate all ultraviolet
problems. We postpone investigation of these issues to future
publications, but let us emphasize here that these issues that
confront us because we have lost the equivalence between
left-invariant, right-invariant and Lorentz invariant measures are
more complex examples of the type of issues that one encounters,
{\it e.g.} when allowing P-parity violation in particle physics
(which actually turns out to be the scenario preferred by {\it
Nature}). The loss of P-parity introduces the arbitrariness
between ``V-A'' and ``V+A'' behaviour which can only be settled by
experiments. In our case besides experiments also the requirement
of mathematical consistency might be useful in identifying the
correct measure. Future more in-depth investigations of this
approach might uncover additional requirements to be satisfied by
the integration measure, thereby reducing the number of choices
available.

Having sketched out our approach to deformed Feynman rules let us
close this Section with some comments on obtaining cross sections
from the amplitudes calculated using the nonAbelian Feynman rules.
This is of course a necessary step since experimental data are
compared to cross sections. The usual formulas cannot be naively
applied in our case since the derivation of cross sections from
amplitudes must now be done consistently with the measurement of
solid angles etc in the noncommutative spacetime. This could be
the most delicate part of our approach because it is the part
where we cannot fully confine the analysis within energy-momentum
space. We indicate here only a general strategy that could be
adopted. First of all, using our principle of normal ordering we
consider normal ordered spacetime expressions as identified with
their classical counterparts for the purposes of specifying solid
angles, etc. Using this identification one is left with the task
of obtaining a consistent deformation of the standard
cross-section formulas. Let us discuss the elements of novelty
required by our framework within the specific example of a
scattering process with two particles in the initial state and two
particles in the final state. The relevant standard cross section
formula in the ordinary commutative Minkowski spacetime is
\begin{eqnarray}
d \sigma \!\! &=& \!\! \frac{d^3 \vec{k}_{f,1}}{ 16 \pi^3 \omega_{f,1}}
\frac{d^3 \vec{k}_{f,2}}{16 \pi^3 \omega_{f,2}}
\frac{|{\cal M} (1_i + 2_i \rightarrow 1_f + 2_f)|^2 }{
|v_{i,1}- v_{i,2}|}
\int \frac{d^3 \vec{q}_{1} }{ 16 \pi^3 \omega_{i,1}}
\frac{d^3 \vec{q}_{2} }{ 16 \pi^3 \omega_{i,2}}
\label{standard}\\
& & ~~~~
|\Phi_{\vec{k}_{i,1}}(q_1)|^2
|\Phi_{\vec{k}_{i,2}}(q_2)|^2
\, 16 \pi^4 ~ \delta^{(4)} \!
\left( (\vec{q}_1,\omega_{i,1})+(\vec{q}_2,\omega_{i,2})-
(\vec{k}_{f,1},\omega_{f,1}) - (\vec{k}_{f,2},\omega_{f,2}) \right) ~,
\nonumber
\end{eqnarray}
where $(\vec{k}_{i,1},\omega_{i,1})$ and
$(\vec{k}_{i,2},\omega_{i,2})$
(respectively $v_{i,1}$ and $v_{i,2}$)
denote energy-momentum (respectively velocity) of the
particles in the initial state,
$d^3k_{f,1}$ and $d^3k_{f,2}$ are infinitesimal volume elements
in the space of momenta of the particles in the final state,
$\Phi_{\vec{k}_{i,1}}(\vec{q})$ and $\Phi_{\vec{k}_{i,2}}(\vec{q})$
are the momentum-space wave functions
of the particles in the initial states, which are assumed to be sharply
peaked around $\vec{q} \sim \vec{k}_{i,1}$
and $\vec{q} \sim \vec{k}_{i,2}$ respectively.

The deformation of the formula (\ref{standard}) requires various
elements of our formalism. Most notably, the energy-momentum
conservation enforced by the $\delta$ function must be implemented
consistently with the nonAbelianess of our energy-momentum space,
and this brings in again some choices with respect to the ordering
of the various momenta entering the sums. The usual problem of
choosing the measure of integration is also present here, but one
would expect this ambiguity to be settled by a requirement of
consistency with the choice of measure adopted for loop-integrals
in Feynman diagrams and in the definition of wave packets. In
particular, at present it appears legitimate to proceed taking
measures according to the left-invariance advocated in Section 2.
Finally the wave functions
$\Phi_{\vec{k}_{i,1}}$,$\Phi_{\vec{k}_{i,2}}$ appear in a very
simple way in equation (\ref{standard}), but this is the result
\cite{bookcross} of the simplicity of the procedure for the
construction of two-particle wave packets in ordinary Minkowski
spacetime. In our case we have already constructed 1-particle wave
packets in Section~2, modulo some possible variations. The usual
definition (as an approximation) for multiple-wave packets is as
the tensor product of 1-particle wave packets, i.e. this in itself
presents no problem in our formalism. E.g.
\eqn{2packet}{ \psi_{a_1,\vec k_1,a_2,\vec k_2}=\psi_{a_1,\vec k_1}\tens
\psi_{a_2,\vec k_2}.}
The ordering of the addition of momenta for in and out states in a
scattering corresponds to the ordering of such tensor products.
For identical particles one could again perform some form of
symmetrization to express our ignorance of which particle should
be on the left and which on the right factor, but in any case
nontrivial implications for the cross-section formula are to be
expected.

\section{Phenomenology}

\subsection{Phenomenology of deformed dispersion relations}

The deformed dispersion relation (\ref{disp}) can have important
implications even though the deformation is only minute (it is
proportional to $\lambda$, which we expect to be close to the
Planck length).
While we derived (\ref{disp}) for scalar particles,
and a rigorous analysis of spin-1 particles
must still await some developments on the mathematics side,
it appears quite plausible that the same dispersion relation,
which is primarily dictated by the deformed symmetries present
in our approach, would also apply to photons.
This would lead to an effect of
energy dependence of the speed of photons
which is large enough for observation in
experiments involving the gamma rays we collect from astrophysical
sources.

In clarifying the origin of this energy-dependence
of the speed of massless particles let us start by observing that
within the stated assumption of existence of
a practical measurement procedure allowing  to treat normal ordered
expressions conventionally it is legitimate
to describe the physical wave velocity
of our noncommutative plane waves (\ref{wave}) according
to the conventional formula
\eqn{vel}{v_i=\frac{d\omega}{d k_i}.}
One may analyse this in terms of a wave packet or, equivalently, by
thinking about one wave (\ref{wave}) at a time. When traveling a
distance $\vec L$ in time $T$ the wave still completes
$n=(\vec k\cdot\vec L+\omega T)/2\pi$ cycles as usual. Hence if we vary
$\vec k$ with the same number of cycles, the arrival time varies
by
\[ \delta T=-\frac{(\vec L+T\vec v)}{\omega}\cdot\delta\vec k\]
as usual, with $v_i$ defined by (\ref{vel}). This is arranged so that
\[ e^{\imath \vec k\cdot\vec x}e^{\imath \omega t}|_{\vec L,T}
=e^{\imath(\vec k+\delta\vec k)\cdot \vec x}
e^{\imath (\omega+\vec v\cdot\delta\vec k)t}|_{\vec L,T+\delta T}\]
when one replaces the noncommutative coordinates $\vec x,t$ by their
measured values as shown. Note that one would obtain quite different answers
due to the noncommutativity of the generators without the normal ordering
assumption for the comparison with measured values. This provides at least
some justification for  (\ref{vel}) within the present framework; a
fuller justification would presumably come out of a more detailed
model of an actual measuring apparatus within a more complete theory.

With this justification, we may combine
(\ref{wave}) and (\ref{disp}) to
finds that the velocity
of massless particles is given by
\eqn{viofe}{v_i = \frac{d \omega}{d k_i}=
\frac{\lambda k_i}{ \lambda^2 {\vec k}^2 +
\frac{\lambda \omega }{ |\lambda \omega |}
\sqrt{\lambda^2 {\vec k}^2} } ~.}
Consequently, the speed of massless particles is given by
\eqn{vofe}{v = \frac{1}{1 + \frac{\lambda \omega}{ |\lambda \omega |}
\sqrt{\lambda^2 {\vec k}^2}}
= e^{- \lambda \omega} \simeq 1 -  \lambda \omega ~,}
where on the right-hand side we expanded for small $\omega$
($\omega \ll \lambda^{-1}$).

This velocity law for massless particles Eq.~(\ref{vofe}) was
already considered in some studies
\cite{kpoin,aln,grbgac} based on the
$\kappa$-Poincar\'e symmetries and studies based on Liouville
non-critical String Theory \cite{grbgac,aemn1}. The fact
that we also encounter this velocity law is of course not
surprising since (in the sense clarified in Section~1) our approach
is consistent with a background $\kappa$-Poincar\'e symmetry.
It is significant however that, thanks to the
 quantum group Fourier transform methods, we could
for the first time discuss corresponding ``plane waves''
(\ref{wave}) and thereby justify Eq.~(\ref{vofe}) as fully
deserving its physical interpretation as velocity law. Instead, in
previous $\kappa$-Poincar\'e approaches this velocity law was only
suggested at a rather heuristic level starting from the properties
of a Casimir and using formal manipulations with generators
$p_\mu$ which, although commuting among themselves, were viewed as
part of a noncommutative deformed Poincar\'e algebra, and with
formulae such as $v_i=d p_0/d p_i$ assumed formally. 
By replacing these
$p_\mu$ by the underlying energy-momentum space with points
$(\vec{k},\omega)$ we are able to compute with the latter, which
are numbers and not formal generators. And we are able to give at
least some justification for (\ref{vel}) through the properties of
the plane waves (\ref{wave}) now at our disposal. Also notice that
our wave equation was not obtained simply using {\em a} Casimir,
which would have not fixed it or its corresponding dispersion
relation uniquely; we also demanded that the dispersion relation
be invariant under group inversion in energy-momentum space.

The velocity law (\ref{vofe}) is a significant prediction
of our proposal since
recent progress in the phenomenology of gamma-ray
bursts \cite{grbnew}
and other astrophysical phenomena \cite{billetal,aemns2}
renders experimentally accessible \cite{grbgac}
the investigation of certain modified velocity laws,
including the ones of type Eq.~(\ref{vofe}).
As explained in Ref.~\cite{grbgac},
these experimental tests are actually rather simple.
In fact,
according to (\ref{vofe}), two signals respectively
of energy $\omega$ and $\omega + \delta \omega$
emitted simultaneously from the same
astrophysical source in travelling a distance $L$
acquire a ``relative time delay'' $|\delta t|$
given by
\begin{eqnarray}
 |\delta t| \sim  {\lambda \, \delta \omega} \frac{L}{c}  ~.
\label{delayt}
\end{eqnarray}
This time delay can be detected if $\delta \omega$
and $L$ are large whilst
the time scale over which the signal exhibits time structure
is small.
These conditions are in particular met by certain
gamma-ray bursts.
We recall that these bursts involve~\cite{review}
typical photon energies
in the range $0.1-100$~MeV
and time structure down to the millisecond
scale has been observed
in the light curves.
According to Eq.~(\ref{delayt})
a signal with millisecond time structure in
a burst of
photons with energies spread over a range of order 10~MeV
coming from a distance of order
$10^{10}$ light years\footnote{The cosmological origin
of at least some GRBs
has been recently established~\cite{grbnew}.}
would be sensitive to $\lambda$ of order
$10^{-35} m \sim L_{p}$.

Already available data~\cite{billetal}
rule out values of  $\lambda$ of order
$10^{-33} m$  in Eq.~(\ref{vofe}),
and planned experiments should achieve
sensitivity to values of $\lambda$
of order $\sim 10^{-35} m \sim L_{p}$ within a few
years~\cite{aemns2}.

Let us also emphasize that in our proposal the ``$v$'' appearing
in Eq.~(\ref{vofe}) is naturally interpreted as the expectation of
a velocity operator in some underlying ``prequantized
pregeometry''. In particular, this implies that a sample of
massless particles of energy $\omega$ would have average speed
given by the $v(\omega)$ of Eq.~(\ref{vofe}) but there would also
be a certain spread $\sigma_v(\omega)$ in the speeds of individual
particles within the sample. Since we do not have any very
definite knowledge
of the structure of the pregeometry quantum system we
are unable to make definite predictions for $\sigma_v(\omega)$,
but we hope experimentalists will find motivation in our analysis
to search for this effect. Additional motivation for this
particular type of experimental investigations comes from
analogous effects encountered in other quantum gravity motivated
studies~\cite{emnnew}.

\subsection{CPT violation}

Another class of recently proposed quantum gravity phenomena have
to do with violations of CPT invariance. This is a rather general
prediction of quantum gravity~\cite{qgcpt}, since most approaches
involve some elements of non-locality (so that one of the
hypotheses of the ``CPT theorem'' does not hold) and/or
decoherence. What is remarkable is that certain quantum gravity
approaches predict violations of CPT invariance large enough to be
detectable by exploiting the properties of the very delicate
neutral-kaon system.

In this Subsection we observe that
the proposal we are putting forward
in the present Article
hosts a mechanism of CPT violation.
Although a detailed study
of CPT violation within our approach
will require the development
of the mathematical tools mentioned in the preceding Section,
we shall also provide here some evidence suggesting that
this CPT violation 
might be tested using the
neutral-kaon system.

The root of CPT violation within our approach
resides in the discretization
of time (in the sense clarified earlier).
Actually, CPT invariance is not necessarily ``lost'':
it can in fact be traded for a novel invariance, which we could see
as a deformed CPT invariance. Our (quantum) deformation of Minkowski
spacetime leads to deformation of $P$ and $T$ transformations.
The situation of CPT transformations in our proposal
is somewhat analogous to the deformations of
Lorentz invariance considered in \cite{kpoin,aln,grbgac},
whose experimental signature would be a violation
of ordinary Lorentz invariance \cite{grbgac}, but
at the fundamental level can be described by replacing the
Lorentz symmetries with a deformed version
of Lorentz symmetries \cite{kpoin,aln}.

In characterizing the deformed CPT invariance
which is consistent with our approach it is important to notice that
in our approach a particle with charges, say, $\alpha,\beta,\gamma$
and momentum $(\vec{k},\omega)$ has as antiparticle
a particle of charges $-\alpha,-\beta,-\gamma$
and momentum not $(-\vec{k},-\omega)$ but
\begin{eqnarray}
(\vec{k},\omega)^{-1}=(-\vec{k}e^{\lambda\omega},-\omega) ~.
\label{neweqone}
\end{eqnarray}
from (\ref{group}). Correspondingly in the loop integrals
of our momentum-space field theory
particles and antiparticles do not
contribute in a totally symmetric way. This is also
 evident when comparing the positive values of energy and
the negative values energy which are consistent with a given
momentum $(\vec{k},\omega)$ according to Eq.~(\ref{disp}) and (\ref{prop}).
That is, {\em if} one takes a usual splitting of momentum into
spatial momentum $\vec{k}$ and energy $\omega$ and carries this over
to the experimental interpretation one can expect to observe the modification
in the group inversion as a breakdown of ordinary CPT invariance.

It may be that such a breaking of ordinary CPT invariance  turns out to be
consistent with quantum mechanics, {\it i.e.} the violations
of CPT invariance may be described as terms in an (effective)
Hamiltonian which governs
otherwise ordinary evolution equations of quantum mechanics.
While this seems rather probable a definite
statement will have to wait more detailed analyses;
in fact, at present one cannot exclude that the novel elements
of our approach (particularly the peculiar nature of time)
could lead to evolution equations not exactly of the type
expected within ordinary quantum mechanics.
We have discussed this possibility already in Section~3.
On the other hand, at present, we are setting up only
a framework for an effective low-energy description of certain
quantum gravity effects and the fact that
the full quantum gravity might require departures from ordinary quantum
mechanics does not necessarily imply that its effective low-energy
descriptions should already incorporate this property.
We emphasize this point because other approaches to quantum gravity
lead to violations of CPT invariance which cannot be
accommodated within the formalism of ordinary quantum
mechanics \cite{ehns,hpcpt,elmncptheory}.

The difference between breaking ordinary CPT invariance
within quantum mechanics \cite{elmncptheory,maianicpt}
and outside quantum mechanics
has been emphasized in work on the neutral-kaon system,
and is accessible experimentally~\cite{elmn}.
The type of breaking of ordinary CPT invariance
which we expect to emerge in future
developments of the approach here proposed
would also be distinguishable from
other proposals because of the fact that here CPT invariance is
replaced by a ``deformed CPT invariance''
whose predictions could be tested experimentally.

\section{Summary and outlook}

The analysis here reported had two objectives which we can now
restate more succinctly using the discussion that preceded.
The first objective was the one of putting on firmer ground
recent ideas on the possibility
that the quantum-group formalism might allow a consistent
formulation of theories with
deformed dispersion relations
of the type which can now be tested \cite{grbgac} using
recent progress in the phenomenology of gamma-ray
bursts \cite{grbnew}
and other astrophysical phenomena \cite{billetal,aemns2}.
In Sections 2, 3 and 6 this more rigorous analysis
was given together with the first elements of a possible
measurement theory for noncommutative spacetimes.
We hope that having established more firmly the possibility
of a consistent formalism for
the mentioned deformed dispersion relations
we will provide additional motivation for experimentalists
to look for this new effect.

Our second objective was to point out the possibility of a new approach
to the construction of field theories on noncommutative spacetimes of
the type here considered (those where a quantum-geometry
transformation\cite{Ma:ista} to a classical but nonAbelian
energy-momentum group is possible), and to discuss some of the issues
 arising. This was done in Section~5.  While several mathematical and
interpretational developments are still required for us to be able to
use this approach for the construction of a meaningful model, we believe
that the procedure here outlined can provide a useful starting point for
future work in this direction. As mentioned in Section~5, it appears
likely that some of these field theories would be well-behaved in the
ultraviolet (they would not require regularization of ultraviolet
divergences). Additional motivation for this research programme should
come from the fact that by constructing (if this indeed turns out to be
possible) a consistent model of particle physics according to the
guidelines described in Section 5 we might then have a formalism which
allows direct/explicit calculation of the mentioned in-vacuo dispersion
effects and CPT-violation effects. The magnitude of these effects could
be related directly and calculably to $\lambda$, while in other
quantum-gravity  formalisms believed
to host these effects the  evaluation of the magnitude of the effects
directly from the original theory turns out to be too difficult (but one
is able to identify in the theory the structures required for the
effects of interest and phenomenological
models \cite{aemn1,emnnew,elmncptheory,emn}
can then be made to
parametrize the magnitude of the effects).

Of course it would also be interesting to
investigate further the idea that our noncommutative
Minkowski spacetime could be used to model some
properties of the ``foamy vacuum'' of quantum gravity.
A natural framework for such studies appears to be provided
by Canonical/Loop quantum gravity \cite{canoloop}.

\vglue 0.6cm
\leftline{\Large {\bf Acknowledgements}}
\vglue 0.4cm

These results were obtained at the Karpacz Winter School in
Polanica, Poland in February 1999; we warmly thank the organisers
for making our collaboration possible. One of us (G.A.C.) would also like to
thank Carlo Rovelli and
Rodolfo Russo for conversations. G.A.C. is
supported by a Marie Curie TMR Fellowship.


\end{document}